# Randomness-based macroscopic Franson-type nonlocal correlation

Byoung S. Ham

Center for Photon Information Processing, School of Electrical Engineering and Computer Science,
Gwangju Institute of Science and Technology
123 Chumdangwagi-ro, Buk-gu, Gwangju 61005, S. Korea
(Submitted on January 16, 2021; bham@gist.ac.kr)

Franson-type nonlocal correlation is a testing tool for Bell inequality violation using noninterfering interferometers, where coincidence measurements involve an interference fringe of $g^{(1)}$ correlation between noninterfering photon pairs. Like the Bell inequality, Franson correlation is also limited to a microscopic regime of entangled photon pairs. Here, randomness-based macroscopic Franson-type nonlocal correlation is presented using polarization-basis coherent superposition of laser light, where probabilistic randomness between bipartite orthonormal bases plays an important role for both Bell inequality and the $g^{(1)}$ correlation. Without contradiction to the conventional understanding of quantumness limited by the particle nature of photons, the proposed Franson correlation can also be extended to a general scheme of macroscopic regimes via coherent superposition.

**Introduction**

Over the last several decades, the Bell inequality [1,2] has been a guideline and a testing tool for the nonlocal quantum correlation in quantum mechanics [3-10]. Franson-type correlation is for an interference version of the Bell inequality based on noninterfering Mach-Zehnder interferometers (MZIs) [11-13], where Franson correlation has been widely adapted for quantum key distribution [14-17]. Like Bell inequality violation, the Franson version involves a typical $g^{(1)}$ amplitude correlation in $g^{(2)}$ intensity correlation via coincidence measurements. Such nonlocal correlation is not compatible with classical hidden variable theory, which is the key concept of the thought experiment proposed by Einstein, Podolsky, and Rosen (EPR) [18]. However, quantum correlation has not been clearly defined with respect to coherence, which has been considered as a classical bound. Here, we present a general model of Franson-type nonlocal correlation compatible with the macroscopic regime for Bell inequality violation with $g^{(1)}$ amplitude correlation. This interpretation does not contradict our conventional understanding limited to a particle nature-based microscopic regime, which cannot be compatible with classical physics due to the uncertainty principle. In that sense, the definition of classical physic in conventional quantum mechanics should be limited to statistical physics without mutual coherence.

Recently, a coherence version of anticorrelation has been proposed [19] to revisit quantumness of the so-called Hong-Ou-Mandel (HOM) dip [20], where anticorrelation is redefined with a definite phase relation between the paired entangled photons. According to the coherence interpretation of anticorrelation [19], the HOM dip is supposed to include $g^{(1)}$ correlation in the coincidence measurements. However, such a $g^{(1)}$ feature has never been observed in any HOM dip experiments [20-24]. Recently, the fundamental physics of no $g^{(1)}$ correlation in a HOM dip has been discussed as a special phenomenon, in which symmetric frequency offsets between paired entangled photons are the source of coherence washout among paired photons generated from spontaneous parametric down conversion (SPDC) processes [25]. Very recently, the coherence version of anticorrelation with a $g^{(1)}$ fringe within a HOM dip has been observed using an attenuated laser by excluding the coherence washout observed in the SPDC case [26]. In addition, another coherence version of a quantum feature [27,28] has also been presented and experimentally demonstrated [29] for a macroscopic version of photonic de Brogle waves [30,31] according to the wave nature of photons, where the conventional emphasis on the particle nature of photons is just a matter of preference in quantum mechanics.

Wave-particle duality is a fundamental law of quantum physics dealing with light and matters [31]. Considering entangled photon pairs generated from SPDC via $\chi^{(2)}$ nonlinear optical processes [32] or $\chi^{(3)}$ four-wave mixing processes [33-34] in an optical medium, the characteristics of entangled photon pairs should cling to their generation process. In general, light is viewed as either a particle or a wave depending on its interpretation or preference of an observer, but cannot be viewed simultaneously in both ways. In other words, the wave nature of entangled photons does not violate quantum mechanics. Thus, the viewpoint of quantum correlation may depend on either a physical interpretation or conceptual understanding, resulting in a matter of preference. In the present work, a clear and general understanding of Franson-type nonlocal correlation is presented to excavate the origin of quantumness. For this, a coherent light source of laser is used instead of probabilistically achieved entangled photons to exclude coherence washout among SPDC-generated photon pairs [25]. To make it conceptually identical to the original scheme of Franson correlation [12,13], a 45 degree-rotated half-wave plate followed by a PBS-based MZI is used to randomly prepare orthogonally polarized photon pairs in each party remotely separately, as happens in a SPDC type-II crystal (see Fig. 1). The random generation of polarized photons provides quantum superposition between two independent bipartite orthonormal bases, where coherence between paired bases plays a key role for quantum correlation, regardless of photon nature in either the particle-based or the wave-based. Specifically, the wave-based approach can expand our scope of quantum correlation into a macroscopic regime via collective control of individual photons or atoms [35,36]. Here, it should be noted that randomness and classicality must be discernible with respect to coherence.

**Analysis**

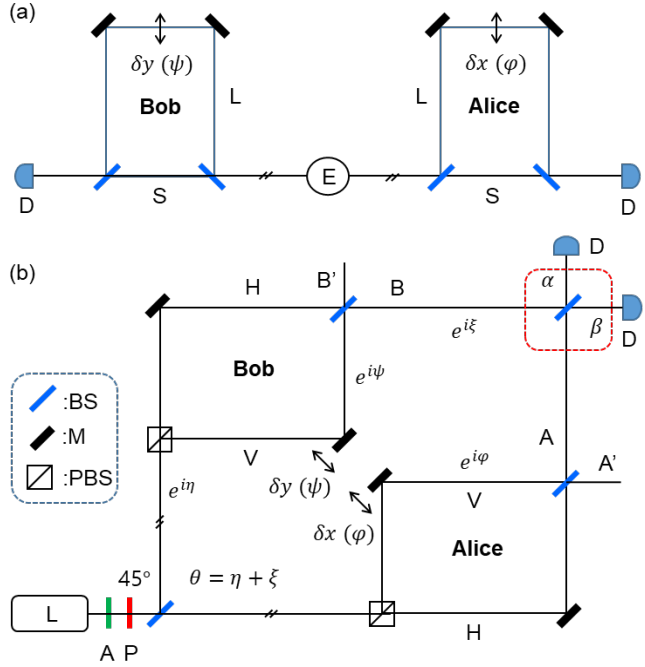

Fig. 1. Schematic of Franson-type nonlocal correlation for (a) original version and (b) coherence version. E: an entangled photon pair, S/L: short/long path, D: detector, L: laser, A: attenuator, P: polarizer, M: mirror, BS: nonpolarizing beam splitter, PBS: polarizing beam splitter, H (V): horizontally (vertically) polarized photon. The 45 degree-rotation of the polarizer P results in random probability in H and V polarizations with respect initially V-polarized light from L. $\varphi = \frac{2\pi}{\lambda}\delta x$ and $\psi = \frac{2\pi}{\lambda}\delta y$, where $\lambda$ is the wavelength of the light from L. The phases η and ξ are due to relative path-length difference between Alice and Bob except for the noninterfering PBS-BS MZI in each side.

Figure 1(a) shows the original scheme of Franson-type nonlocal correlation [12,13], and Fig. 1(b) shows the corresponding coherence version discussed herein. The noninterfering MZI in Fig. 1(a) is mimicked by a PBS-BS composed noninterfering MZI in Fig. 1(b), where the short-short and long-long path superposition in Fig. 1(a) is replaced by polarization superposition between horizontal (H) and vertical (H) ones. For proof of principle, a typical $g^{(1)}$ correlation scheme with a BS is also added as shown in the red-dotted box in addition to the coincidence measurement without it. Here, it should be noted that coincidence measurement in Fig. 1(a) is a way to trace the nonlocal correlation between coupled photons via quantum superposition principle. Ragarding the particle nature of paired photons in a coupled system, Heisenberg's uncertainty principle has nothing to do with the details of their relative phase information. For an implementation purpose of the nonlocal correlation such as for quantum key distribution [14-17], however, the cross-interference scheme of the red dotted box can simply be replaced by a self-interference scheme relying on the source L as usual. Here, it should be emphasized that Fig. 1(b) does not have to be limited to the conventional particle nature of photons, but can be extended into a macroscopic regime with the wave nature of photons via collective control of ensemble photons [19,25-29].

Starting from a coherent ensemble of photons generated from the laser L in Fig. 1(b), a neutral density filter (A) is inserted right after the source L to reconfigure the coherent scheme into a microscopic one composed of doubly-bunched photon pairs governed by Poisson statistics. The double-to-single photon ratio is determined by the mean photon number $\langle n \rangle$. For $\langle n \rangle = 0.012$, the double-to-single photon ratio in the attenuated scheme of Fig. 1(b) has been experimentally demonstrated to be 0.005 [26]. Each photon pair is split into two paths for Alice and Bob equally by a BS, where each photon's polarization is randomly provided by a 45 degree-rotated half-wave plate (P). This random polarization of H (horizontal) and V (vertical) in Fig. 1(b) is similar but not identical to the SPDC type-II case, in terms of entanglement [6-10]. Single photons in Fig. 1(b) do not contribute to the coincidence measurements [26]. Three or more bunched photons are also neglected due to Poisson statistics [26]. The coherence length of the photon pair is predetermined by the laser L, which spans from a few millimeters to a few kilometers depending on the commercially available laser systems.

The type-II SPDC-based Bell inequality like H-V polarization-based Franson correlation is analyzed using the polarization state superposition in a PBS-BS MZI via the 45 degree-rotated half-wave plate in Fig. 1(b):

$$|\Psi\rangle_A = (|H\rangle_A + ie^{i\varphi}|V\rangle_A)/\sqrt{2}, \qquad (1)$$
$$|\Psi\rangle_B = ie^{i\theta}(|H\rangle_B + ie^{i\psi}|V\rangle_B)/\sqrt{2}. \qquad (2)$$

where the amplitude of H and V is equivalent to $E_0$, and the term $ie^{i\theta}$ is due to the total relative path-length difference ($\theta = \eta + \xi$) between Alice and Bob. For coincidence detection, the last BS (see the red-dotted box) for α and β interference (see the red-dotted box) is removed to satisfy the nonlocal correlation between Alice and Bob. The subscript A and B in equations (1) and (2) indicate Alice and Bob, respectively. Due to the Fresnel-Arago law [37], however, H-V interference terms interfering on a BS are automatically removed [38]. Here, the relative phases φ and ψ are related with the path-length (δx; δy) difference inside the PBS-BS MZI (small squares for Alice and Bob), which are precisely controlled without violation of the uncertainty principle.

For analysis of the coherence version of Franson correlation in Fig. 1(b), we first show single photon-based coincidence measurements and then expand it for a coherent ensemble. The analytic expression for the coincidence detection $R_{AB}$ measured by both detectors remotely located in Alice's and Bob's sides is directly obtained from equations (1) and (2):

$$\begin{aligned}
R_{AB} &= \tfrac{1}{4}\langle (\langle H|_A - ie^{-i\varphi}\langle V|_A)(\langle H|_B - ie^{-i\psi}\langle V|_B)(|H\rangle_A + ie^{i\varphi}|V\rangle_A)(|H\rangle_B + ie^{i\psi}|V\rangle_B)\rangle \\
&= \tfrac{1}{4}\langle (\langle H|H\rangle_{AB} + e^{-i(\varphi-\psi)}\langle V|V\rangle_{AB})(\langle H|H\rangle_{BA} + e^{i(\varphi-\psi)}\langle V|V\rangle_{BA})\rangle \\
&= \tfrac{I_0^2}{2}\langle 1 + \cos(\varphi - \psi)\rangle, \qquad (3)
\end{aligned}$$

where $I_0 = E_0 E_0^*$, $\varphi = \tfrac{2\pi}{\lambda}\delta x$, and $\psi = \tfrac{2\pi}{\lambda}\delta y$. Thus, equation (3) shows the same result as the original Franson-type nonlocal correlation [12,13]. With precise control of φ and ψ, the coincidence measurement in equation (3) shows a definite Bell inequality violation with a visibility greater than $1/\sqrt{2}$ [1-17].

On the other hand, the coherence version of Franson-type nonlocal correlation in Fig. 1(b) can be described for the interference of coherent fields A ($\Psi_A$) and B ($\Psi_A$) as follows:

$$\begin{aligned}
I_\alpha &= \tfrac{1}{4}\{[(H_A + ie^{i\varphi}V_A - e^{i\theta}(H_B + ie^{i\psi}V_B))][(H_A^* - ie^{-i\varphi}V_A^* - e^{-i\theta}(H_B^* - ie^{-i\psi}V_B^*))]\} \\
&= \tfrac{1}{4}[H_A H_A^* + V_A V_A^* + H_B H_B^* + V_B V_B^* - (H_A H_B^* + V_A V_B^* e^{i(\varphi-\psi)})e^{-i\theta} - (H_B H_A^* + V_B V_A^* e^{-i(\varphi-\psi)})e^{i\theta}] \\
&= \tfrac{I_0}{2}[2 - \cos(\theta) - \cos(\varphi - \psi - \theta)], \qquad (4)
\end{aligned}$$

where $V_i$ and $H_i$ indicate polarized fields in the Alice's and Bob's sides, whose amplitude is $E_0$. All $V_i H_j$ interference terms are deleted due to the Fresenl-Arago law [37]. Each output amplitude superposed by equations (1) and (2) is as follows: $E_\alpha = \tfrac{1}{\sqrt{2}}(\Psi_A + ie^{i\xi}\Psi_B)$; $E_\beta = \tfrac{i}{\sqrt{2}}(\Psi_A - ie^{i\xi}\Psi_B)$; $I_j$ is defined as $I_j = E_j E_j^*$. In the view point of coherence optics, Fig. 1(b) satisfies a general MZI scheme (see the big square including both noninterfering PBS-BS MZIs). For coherence optics of MZI, $E_0$ can be either a single photon [39] or coherent light. Likewise,

$$I_\beta = \frac{I_0}{2}[2 + \cos(\theta) + \cos(\varphi - \psi - \theta)]. \tag{5}$$

Although equations (4) and (5) are different from equation (3), they share the $g^{(1)}$ correlation feature (see Fig. 2) [6,12,38]. Unlike coincidence measurements for single photons, the coherence version of Franson-type nonlocal correlation in equations (4) and (5) includes both classical and quantum limits depending on the relative phase between $\varphi$ and $\psi$ as well as $\theta$. In other words, the quantum feature of the coherence version of Franson correlation relies on the relative phase between the coherent photon pair for polarization basis superposition. The $\theta$ effect resulting from the BS-BS MZI (see the big square in Fig. 1(b)) is viewed for a HOM dip [19]. Thus, Franson correlation in Fig. 1(b) has a chance to be compared with an interference model of HOM dip for the fundamental physics of quantumness. Here, it should be noted that the coincidence measurement has no practical advantages compared with the wave nature of photons because wavelength-limited $g^{(1)}$ is much more sensitive than coherence length-limited $g^{(2)}$. This is why $g^{(1)}$ correlation should appear within $g^{(2)}$ correlation in principle.

Figure 2 shows numerical simulations for equations (3)-(5), where a single-shot near perfect measurement is another impotant benefit in the coherence version in addition to the macroscopic feature. Figures 2(a) and (b) are for single photon-level coincidence measurements based on equation (3) as a reference, whereas Figs. 2(c) and (d) are for the coherence counterparts based on equations (4) and (5). In terms of coincidence measurements for the particle nature of photons, the relative phase-dependent $R_{AB}$ in Fig. 2(b) shows a defnite coherence feature of $g^{(1)}$ correlation as observed in ref. 12. If there is no superposition between polarization bases via the noninterfering PBS-BS MZIs, then the $g^{(1)}$ correlation term in equation (3) dispappers, resulting in only the $\tau$ dependence of $g^{(2)}(\tau)$:

$$R_{AB} = \langle(-ie^{-i\varphi}\langle V|_A)(-ie^{-i\psi}\langle V|_B)(ie^{i\varphi}|V\rangle_A)(ie^{i\psi}|V\rangle_B)\rangle = I_0^2. \tag{6}$$

Thus, the observed $g^{(1)}$ feature in the original Franson experiment [12] is rooted in the superposition between short-short and long-long path propagating photon pair. In other words, the quantum feature of Franson correlation is due to the coherence between paired photons via random basis superposition (discussed below).

The output intensities of $I_\alpha$ and $I_\beta$ in Fig. 1(b) are opposite each other, satisfying general MZI physics as shown in Figs. 2(c) and (d). The maxima of $I_\alpha$ occur at $\psi + \theta = \varphi \pm \pi$ to satisfy MZI physics as mentioned above, whereas $I_\beta$ at $\psi + \theta = \varphi$. Figures 2(e) and (f) represent output intensities as a function of $\varphi$ for different $\theta$. As expected, the output intensities behave oppositely, but swing across the half-sum value $I_0$ as $\theta$ varies as indicated by an arrow. Although the global phase control of $\theta$ in one path (Bob) with resespct to another (Alice) has no effect on the direct coinicidence measurement of equation (3) [1-17], it affects the quantum feature as described by anticorrelation (see also Fig. S1. of the Supplementary Informaton) [19]. For this, both intensities $I_\alpha$ and $I_\beta$ in Fig. 2(e) and (f) are multiplied as a function of $\theta$ for different values of $\psi$ for a fixed $\varphi = 0$ (see also Fig. S2 of the Supplementary Information). If the anticorrelation condition is satisfied as shown in the blue curve at $\theta = \pm n\pi$ for $\varphi = \psi = 0$, the effect of photon bunching occurs at every $\theta = \pm n\pi$ [19,24,25]. Obviously, the degraded anticorrelation such as in the red cureve ($\psi = \frac{\pi}{2}$; $\varphi = 0$) can be fixed if $\varphi$ can be compensated to be $\varphi = \psi$ in the case of $\theta = n\pi$ (see Fig. 2(h) and bottom pannels of Fig. S2 of the Supplementary Information). However, there is no way to fix the correlation degradation if $\varphi \neq \psi$ ($\varphi \neq \psi \pm \pi$). As a result, the main parameters for Franson correlation are $\varphi$ and $\psi$. This is the fundamental physics of Franson-type nonlocal correaltion how it can be used for quantum key distribution [14-17].

Figure 2(h) shows $I_\alpha$ versus $\theta$ for different $\varphi$ with respect to a fexed $\psi = 0$, where the phase $\varphi$ increases from 0 (blue curve) toward $2\pi$ as indicated by the increasing numbers (color matched). As analyzed in Fig. 2(g), the loss of anticorrelation or degradation of Franson correlation is due to the breakage of the anticorrelation condition ($\varphi = \psi$; $\varphi = \psi \pm \pi$). This shows the nonlocal phase correlation between $\psi$ (Bob) and $\varphi$ (Alice). If $\theta (= \eta + \xi)$ is shifted by $\pi/2$, then either Bob's phase $\psi$ or Alice's $\varphi$ needs to be compensated by the same amount of shift for the condition $\varphi = \psi + \theta$ (see Fig. S3 of the Supplementary Information). However, the $\theta$ compensation cannot retrieve the visibility fully unless the anticorrelation condition between $\varphi$ and $\psi$ is kept as shown in Fig. 2(h). This means that the quantumness is mainly related with the randomness-based superposition between two noninterfering PBS-BS MZIs via nonlocal correlation, where the nonlocal correlation is rooted in the random-basis superposition via phase coherence. By the way, another set of PBS-BS MZI outputs A' and B' also results in the same features as in equations (3)-(6) (see Figs. S4 and S5 of the Supplementary Information): $I_{\alpha'} = I_\alpha$; $I_{\beta'} = I_\beta$.

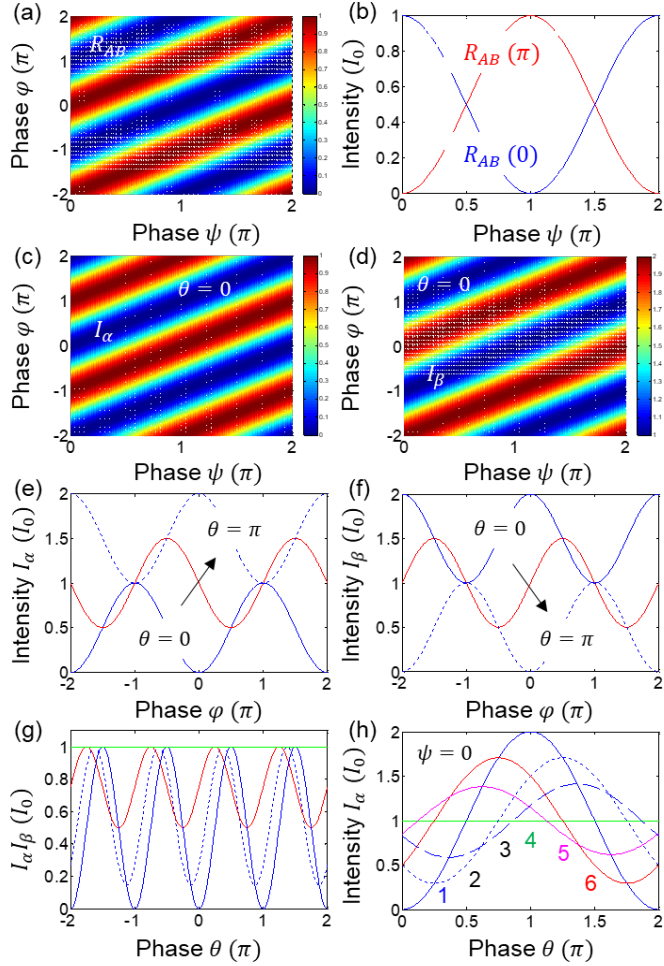

Fig. 2. Numerical simulations for equations (3)-(5). (a) and (b) Normalized $R_{AB}$ for $\theta = 0$. The values in parenthesis are for $\varphi$. (c) and (d) Intensities $I_\alpha$ and $I_\beta$ for $\theta = 0$. (e) and (f) Intensities $I_\alpha$ and $I_\beta$ for $\theta = 0$ (blule), $\theta = \pi/2$ (red), $\theta = \pi$ (dotted). (g) Intensity product $I_\alpha I_\beta$ versus phase $\theta$ for $\varphi = 0$ with $\psi = 0$ (blue), $\psi = \pi/4$ (dotted), $\psi = \pi/2$ (red), and $\psi = \pi$ (green). (h) $I_\alpha$ for $\varphi = 0$ (blue); $\varphi = \pi/2$ (dotted); $\varphi = 3\pi/4$ (dashed); $\varphi = \pi$ (green); $\varphi = 5\pi/4$ (magenta); $\varphi = 3\pi/2$ (red).

**Conclusion**

In conclusion, the origin of entanglement or nonlocal correlation observed in Franson-type nonlocal correlation was investigated using a classical model of random polarization bases. To make the classical model similar to the original Franson correlation, random generation of polarized photons or light fields was provided for superposition in a PBS-BS noninterfering MZI at each party, which is equivalent to the original SPDC-based scheme. From the analytical approach, the first conclusion regarding the origin of Franson correlation was found in the random polarization-basis superposition. From numerical simulations, the second conclusion was found in the definite phase relationship between superposed photons or fields to satisfy $g^{(1)}$ correlation. Thirdly, the analytically obtained Franson correlation in the classical model was tested for anticorrealtion, the so-called Hong-Ou-Mandel dip. From this, the two remotely separated coherently coupled output fields were made to be interfered on a BS. As a result, the same phase relationship achieved in ref. 19 was demonstrated for the bunching effects of anticorrelation. This phase relationship is between two noninterfering MZIs remotely separated. Like conventional understanding of nonlocal correlation based on random-basis superposition in SPDC generated photon pairs, same results were obtained classically and macroscopically in the present classical model. Thus, the fundamental question of quantumness in nonlocal correlation is now asked again for the correct definition and better understanding of the quantum features. Such an answer may lead us to coherence quantum information for future deterministic and macroscopic quantum technologies.

Acknowledgment
BSH acknowledges that the present research was supported by GIST via GRI 2021.